\documentclass{aa}  

\usepackage{graphicx}
\usepackage{txfonts}
\usepackage{multirow}
\usepackage{hyperref}
\begin{document}

   \title{On the Origin of the Globular Cluster FSR 1758}
   
\author{Fu-Chi Yeh
            \inst{1} 
            \and
            Giovanni Carraro
            \inst{1}
              \and
            Vladimir I. Korchagin
            \inst{2}
            \and
            Camilla Pianta
            \inst{1}
            \and
            Sergio Ortolani
            \inst{1}
          }
            
\institute{Department of Physics and Astronomy {\it Galileo Galilei}, Vicolo Osservatorio 3, I-35122, Padova, Italy\\
              \email{giovanni.carraro@unipd.it}
              \and
              Southern Federal University, Rostov on Don, Russian Federation
              }

     \date{Received  ; accepted  }

 
  \abstract
   {Globular clusters in the Milky Way are thought to have either an {\it in situ} origin, or to have been deposited in the Galaxy by past accretion events, like the spectacular Sagittarius dwarf galaxy merger.}
   {We aim to probe the origin of the recently discovered globular cluster FSR 1758, often associated with some past merger event, and which happens to be projected toward the Galactic bulge,  
   by a detailed study of its Galactic orbit, and to assign it to the most suitable Galactic component.}
   {We employ  three different analytical time-independent potential models to calculate the orbit of the cluster by using the Gauss Radau spacings integration method.
    In addition, a time-dependent bar potential model is added to account for the influence of the Galactic bar.  We run a large suite of simulations to account for the uncertainties in the initial conditions, in a Montecarlo fashion.}
   {We confirm previous indications that the globular cluster FSR 1758  possesses a retrograde orbits with high eccentricity. The comparative analysis of the orbital parameters of star clusters in the Milky Way, in tandem with recent metallicity estimates, allows us to conclude that
   FSR1758 is indeed a Galactic bulge  intruder.  The cluster can therefore be considered an old metal poor halo globular cluster formed {\it in situ}  and which is passing right now in the bulge region.  Its properties, however, can be roughly accounted for also assuming that the cluster is  part of some stream of extra-Galactic origin.}
   {We conclude that assessing the origin, either Galactic or extra-galactic, of globular clusters is surely a tantalising task. In any case, by using an {\it Occam's razor} argument, we tend to prefer an {\it in  situ} origin for FSR 1758.}


   \maketitle
%

\section{Introduction}

Being  the oldest stellar systems in the Galaxy, globular clusters (GCs) have been intensely studied  in their spatial distribution, kinematic properties, and chemical composition to probe the assembly history of the Galaxy. They can be classified into three subsystems: the bulge/bar, the old halo and the young halo systems \citep{Zinn1985, Zinn1993, Minniti, Cote1999}. The old halo is thought to have formed  from the halo collapse that happened before the formation of the Galactic disk, while the young halo GCs would be remnants of accretion events of satellite dwarf galaxies into the Milky Way.
On the other hand, the bulge/bar typilcally system form out of instabilities in a nearly-in-equilibrium rotating disk  immersed in a dark matter halo \citep{Com1981,Port2017}.
Therefore, assessing the origin of individual globular clusters allows us to cast light on the evolutionary history of the Milky Way. 
    
FSR 1758 (\cite{FSR07}) has been recently discovered to be a  globular cluster presently located near the Galactic bulge.  It was first studied  by \cite{Barba} who question whether the object is a typical metal-poor GC residing in the Milky Way or the core of an accreted dwarf galaxy.The latter scenario was supported by the common proper motions in the surrounding halo stars which could indeed  be the tidal debris of the dwarf galaxy. Soon after, \cite{Simpson} argued that the halo stars are in fact not associated with the cluster because of the distinct distributions in proper motion, colour and parallax between the cluster members and the halo stars. \citeauthor{Simpson} also back-integrated the orbit of FSR 1578 for 2.5 Gyr. He suggested  that FSR 1758 is most probably a genuine MW GC. However, he also added that because of  the lack of solid estimates  of radial velocity and metallicity it is not possible to firmly exclude an accretion origin for FSR 1758. 
More recently, \cite{Myeong2019} suggested that FSR 1758 is a probable member of Sequoia based on the distribution of Galactic GCs in the action spaces. Finally, very recently \cite{Villanova} obtained the first high-resolution metallicity measurement, which they combined with a new, but simple, orbit calculation. They did not find any metallicity spread and a significant Na-O anti-correlation, as expected for globular clusters.
However, they favor the conclusion that FSR 1758 is genuine member of the Sequoia merger event.

Clearly, assessing the origin and parent stellar population of FSR 1758 is a difficult task, which requires both high quality observational data and a more comprehensive theoretical study of the cluster orbit.
In this paper, we pursue the second avenue, and investigate the orbit of FSR 1758  in a statistical way by employing  three different Galactic potential models which also include the Galactic bar. This in fact is expected to play a major role
in shaping the cluster orbit, giving its present day location,

The layout of the paper is as follows. In Sect.~2 we describe how orbits are computed, while in Sect~3 the results of orbit calculations is discussed, and the outcome is compared with stars clusters in the Milky Way in Sect.~4.

\section{Orbital Calculation}

 We compute  the orbit of FSR 1758 by integrating the equation of motions 1.25 Gyr backward in time  with Gauss-Radau spacings of 15th order (GR15, \citealt{RK15}), employing the  three different potential models described in \citet{Irrgang}.  The time integration does not correspond to the cluster age. This choice is motivated by the fact that (1) integrating longer in time does not have much physical meaning, since the Galactic potential is expected to have changed over the Galaxy lifetime, and also because  (2) we are interested in the actual orbital parameters. 
We anyway tested a longer integration time of 5 Gyrs. To this aim we run a set of simulations using Model 1, and analysed the output in the same way as for the 1.25 Gry simulations. We obtained the following results:  $\Delta R_{peri} = 0.01$ kpc, $\Delta R_{apo} = 0.2$ kpc, $\Delta z_{max} = 0.05$ kpc, $\Delta e = 0.03 $, $\Delta E = 0.06 $  $(\frac{kpc}{Myr})^2$, and $\Delta L_{z} = 0.03 $  $\frac{kpc^2}{Myr}$.
 
 Finally, we ignore here dynamical friction. This is quite an acceptable choice for a cluster moving fast in a highly eccentric orbit. If we use \citep{Bin2008} formula
 adopting the most accepted values for FSR1758 mass and distance we obtain $t_{friction} = 8.6 \times 10^{9} yr$. Being this derived for a circular orbit,
we can infer that the real dynamical friction time would be much larger, spending the cluster most of the time in the low density halo regions.

The initial conditions of the GC FSR 1758 are provided by \cite{Villanova} where high dispersion spectra are discussed to derive an  accurate radial velocity, while proper motions are taken from Gaia DR2. Initial conditions read:
($\alpha$, $\delta$, d, $\mu_{\alpha}$, $\mu_{\delta}$, $v_r$) = (262.81$^{\text{o}}$, -39.82$^{\text{o}}$, 11.5$\pm$ 1.0 kpc, -2.79$\pm$ 0.0097mas/yr, 2.6$\pm$ 0.009 mas/yr, 226.8$\pm$ 1.6 km/s). 

The corresponding positions and velocities in a Galacto-centric coordinate reference system are listed in Tab.\ref{tab:outputs} adopting  a distance of the Sun to the Galactic Center: $R_{\odot}=8.2\pm 0.1$ kpc, a solar offset from local disk: $z_{\odot}=25 \pm 5$ pc, and a Sun's tangential velocity relative to Sgr A*: $V_{g,\odot}=248\pm$ 3 km/s according to \cite{Bland-Hawthorn2016}. 
In generating initial conditions we followed \cite{Jon87} closely.  Uncertainties are derived by using Monte Carlo simulations. In details, we  start from the observational initial values (means and associated errors, assumed normal) and extract a random value inside the range defined by 2$\times \sigma$.  This procedure was repeated 100 times with the aim of providing statistical estimates for the orbital parameters. These were in turn derived as means and associated errors of the 100 simulations.

\begin{table*}
\centering
\setlength\tabcolsep{2pt}
\begin{tabular}{|c|c|c|c|c|c|c|}
\hline
Inputs & X (kpc) & Y (kpc) & Z (kpc)& U (km/s)& V (km/s)& W (km/s)\\ \hline
-- & 2.87$\pm$0.97 & -2.14$\pm$0.18 & -0.66$\pm$0.06 & 252.67$\pm$2.24  & 245.55$\pm$2.88 & 198.63$\pm$17.68 \\ \hline  
 
Outputs & $R_{peri}$(kpc) & $R_{apo}$(kpc) & $Z_{max}$(kpc) & e & E (kpc/Myr)$^2$ & $L_z$(kpc$^2$/Myr) \\ \hline
Model I & 3.59$\pm$0.8 & 14.98$\pm$3.53 & 6.3$\pm$2.06 & 0.6$\pm$0.02 & -0.14$\pm$0.014 & -1.27$\pm$0.3 \\ \hline 
Model II & 3.65$\pm$0.8 & 16.75$\pm$5.75 & 7.34$\pm$3.2 & 0.63$\pm$0.03 & -0.13$\pm$0.016 & -1.3$\pm$0.29 \\ \hline
Model III & 3.81$\pm$0.86 & 18.9$\pm$4.45 & 8.33$\pm$3.05 & 0.66$\pm$0.02   & -0.29$\pm$0.01 & -1.34$\pm$0.3 \\ \hline
\end{tabular}
\caption[Orbital parameters derived from three models ]{The first two rows list the initial conditions, whilst the  last four show the corresponding calculated orbital parameters in three different potential models assuming for the bar component $\Omega=41$ kpc$^{-1}$ km/s. }
    \label{tab:outputs}
\end{table*}
 
Each model is composed of three time-independent, axisymmetric components: a central bulge, a flat disk and a spherical halo. The bulge and the disk potentials have the same form in three models, while the halo varies.  In addition, the time-dependent non-axisymmetric bar potential was also considered to investigate its influence on the orbit. When adding the bar potential, we assumed that the mass of the bulge is transferred immediately to the mass of the bar at the epoch of bar formation, deep in the past. Since we  do not know precisely when and how the bar formed,  for our purposes of deriving actual orbital parameters the bar potential is an additive fixed (except for the time dependence) term. 
The pattern speeds of the bar applied here are $\Omega=41, 50, 60$ kpc$^{-1}$ km/s. These values are taken from \citep{San2019}, \citep{Min2007}, and \citep{Deb2002}, respectively. Clearly, when the bar potential is included, conservation of energy and angular momentum are not guaranteed. 

 The form of three potential models, as well as the bar potential, are listed below. \newline
-- The potential of the bulge:
\begin{equation}
\label{bulge}
    \Phi_b(R)  = - \dfrac{M_b}{\sqrt{R^2+b_b^2}}
\end{equation}
-- The potential of the disk:
\begin{equation}
\label{disk}
    \Phi_d(r,z) = - \dfrac{M_d}{\sqrt{r^2+(a_d + \sqrt{z^2+b_d^2})^2}}
\end{equation}
-- The halo potential of Model I ($\gamma =2$):
\begin{equation}
\label{halo1}
    \Phi_h(R)  = \begin{cases}
    \dfrac{M_h}{a_h}\Bigg(\dfrac{1}{(\gamma -1)}ln \bigg(\dfrac{1+(\dfrac{R}{a_h})^{\gamma -1}}{1+(\dfrac{\Lambda}{a_h})^{\gamma -1}}\bigg) - \dfrac{(\dfrac{\Lambda}{a_h})^{\gamma -1}}{1+(\dfrac{\Lambda}{a_h})^{\gamma -1}}\Bigg)&\text{, if } R < \Lambda.    \\
    \dfrac{M_h}{R}\dfrac{(\dfrac{\Lambda}{a_h})^{\gamma}}{1+(\dfrac{\Lambda}{a_h})^{\gamma-1}} &\text{, if } R > \Lambda.
    \end{cases}
\end{equation}
-- The halo potential of Model II:
\begin{equation}
\label{halo2}
    \Phi_h(R) = - \dfrac{M_h}{a_h}\ln \bigg( \dfrac{\sqrt{R^2+a_h^2}}{R}    \bigg)
\end{equation} 
-- The halo potential of Model III:
\begin{equation}
\label{halo3}
    \Phi_h(R) = -\dfrac{M_h}{R}\ln \bigg( 1+\dfrac{R}{a_h} \bigg)
\end{equation}
$M_b$, $M_d$ and $M_h$ represent the total mass of the bulge, disk and the halo. The $\Lambda$ symbol in Eq.\eqref{halo1} is a cut-off radius to avoid an infinite halo mass. The parameters $b_b$, $a_d$  and $a_h$ control the scales of bulge, disk and halo component. The value of $b_d$ adjusts the scale height of disk.

As for the bar, we use Ferrers model and  the density $\rho(x, y, z)$  is given by

\begin{align}
\label{bar_density}
\rho(x,y,z) = \begin{cases}
\rho_c (1-m^2)^2&\text{, if m} < 1, \\
0 &\text{, if m} > 1,
\end{cases}
\end{align}
where $\rho_c=\dfrac{105}{32\pi}\dfrac{GM_{bar}}{abc}$, $M_{bar}$ is the total mass of the bar transferred from the mass of the bulge, and $m=\dfrac{x^2}{a^2}+\dfrac{y^2}{b^2}+\dfrac{z^2}{c^2}$.  According to  \citet{Pichardo}, the major axis half-length $a = 3.14$ kpc, and the axial ratio $a:b:c = 10:3.75:2.56$.  The present position angle of the longest axis of the bar with respect to the line of sight is $25^{\text{o}}$ as in the recent results of  \cite{Bovy}. 

According to Chandrasekar (1969, p.53), the potential of bar in the form of Eq.\eqref{bar_density} is expressed as:
\begin{equation}
\Phi = -\pi G abc \dfrac{\rho_c}{n+1} \int_{\lambda}^{\infty}\dfrac{du}{\Delta(u)}(1-m^2(u))^3 \text{, where } 
\end{equation}
\begin{equation}
 m^2(u)=\dfrac{x^2}{a^2+u}+\dfrac{y^2}{b^2+u}+\dfrac{z^2}{c^2+u} \text{, and }
 \end{equation}
 \begin{equation}
 \Delta^2(u)=(a^2+u)(b^2+u)(c^2+u).
\end{equation} 
$\lambda$ is the positive solution of $m^2(\lambda)=1$ such that outside the bar $\Phi=0$. Inside the bar $\lambda=0$.

All these models must be constrained by some observational data  in order to make sure that the total analytic Galactic potential resembles the real Galaxy. These constraints are the Galactic rotational curve, the local mass density,  and the local surface density. They can be derived using the following equations: 

\begin{equation}
\label{vc}
v_c = \sqrt{r^{'} \dfrac{d\Phi(r, 0)}{dr^{'}} } \bigg{|} _{r^{'}=r}
\end{equation}

\begin{equation}
\label{density}
\rho_{\odot}  = \rho_b(r_{\odot}) +  \rho_d(r_{\odot}) +  \rho_h(r_{\odot}) 
\end{equation}

\begin{equation}
\label{sd}
\sum_{1.1}  = \int_{-1.1kpc}^{1.1kpc} \big{[} \rho_b(r_{\odot},z) +  \rho_d(r_{\odot},z) +  \rho_h(r_{\odot},z) \big{]} dz
\end{equation}

 \cite{Holmberg2000} and \cite{Holmberg2004} derived the local density of disk  to be $\rho_{\odot}=0.102\pm 0.010 \text{M}_{\odot}\text{pc}^{-3}$ using Hipparcos data on a volume-complete sample of A and F stars, and the surface density  to be $\sum_{1.1}=74\pm 6\text{M}_{\odot}\text{pc}^{-2}$ from K-giant stars at the SGP. 
    
 The parameters of three potential models are found by fitting the derived constraints to the observed values by means of  $\chi^2$ minimisation. They are listed in Tab.\ref{tab:model01}, Tab.\ref{tab:model02}, and Tab.\ref{tab:model03} for Model I, II,  and III, respectively.  The obtained parameters are compatible with  \citep{Irrgang} models within the uncertainties.  
 We are aware that in the inner regions of the Galaxy the rotation curve is poorly constrained \citep{Cher}, and this affects the vast majority of potential models available in the literature. In the present study, the orbits we computed do not bring FSR 1758 inside the bar, and therefore the cluster spends the majority of its lifetime outside this critical region.

\begin{figure*}
    \centering
    \includegraphics[width=\textwidth]{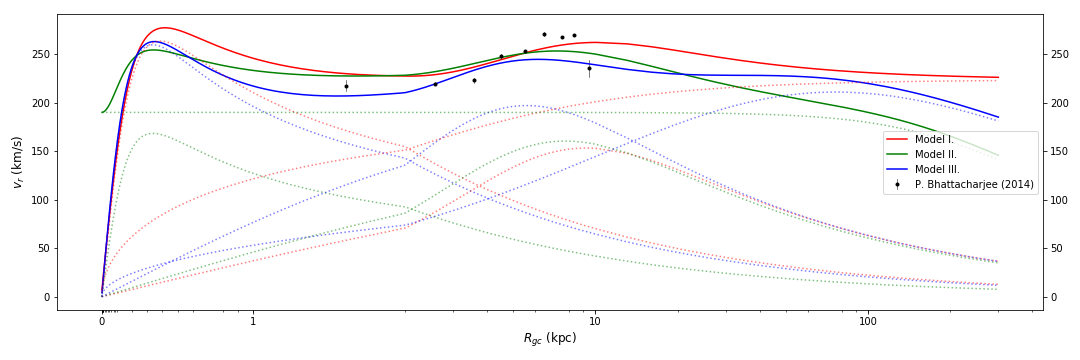}
    \caption[The galactic rotational curve.]{The galactic rotational curve. The colours in red, green, blue represent respectively potential of model I, model II,  and model III. For each model, the contribution of galactic components: bulge contributes the most within 2 kpc, the disk reached a peak at around 10 kpc, and the extended rotational curve is due to halo component. Observational data from \cite{vc} within 10 kpc is plotted in black with the error lines.}
    \label{fig:vc}
\end{figure*}

\begin{table*}
    \centering
    \begin{tabular}{|cccc|}
    \hline  
      Parameters & Value$^*$     & Best fit  & Derived \\ \hline
        $M_b(10^{10}\text{M}_{\odot})$     & 0.950925  & 1.098 & 1.098 \\
        $M_d(10^{10}\text{M}_{\odot})$    & 6.6402    & 8.9 & 6.497\\
        $M_h(10^{10}\text{M}_{\odot})$    & 2.36685    & 2.6657 & 2.15\\ 
        $b_b$(kpc)    & 0.23     & 0.27 & {}\\
        $a_d$(kpc)    & 4.22     & 6.22 & {}\\
        $b_d$(kpc)    & 0.292     & 0.33 & {}\\
        $a_h$(kpc)    & 2.562     & 2.39 & {}\\ 
        $\chi^2$ &    --   &  1.03  & {} \\ \hline \hline
     Constraints & Observed  & Best Fit & {}\\ \hline
        $V_r$    & see \cite{vc} & see Fig.\ref{fig:vc} & {}\\
        $\rho_{\odot}$ & 0.102$\pm$0.01 & 0.128 & {} \\
        $\sum_{1.1}$  & 74$\pm$6        & 74.5 & {}\\ \hline
    \end{tabular}
    \caption[Parameters of potential Model  I]{Parameters of Model I. $^*$ The values are extracted from Table 1 of \cite{Irrgang}. The best fit values of parameters are obtained via $\chi^2$ minimisation. }
    \label{tab:model01}
\end{table*}
\begin{table*}
    \centering
    \begin{tabular}{|cccc|}
    \hline 
      Parameters & Value$^*$     & Best fit  & Derived \\ \hline
        $M_b(10^{10}\text{M}_{\odot})$     & 0.406875  & 0.3 & 0.39\\
        $M_d(10^{10}\text{M}_{\odot})$    & 6.577425    & 8.16 & 7.945\\
        $M_h(10^{10}\text{M}_{\odot})$    & 162.110625    & 160.14 & 71.865\\ 
        $b_b$(kpc)    & 0.184     & 0.238 & {}\\
        $a_d$(kpc)    & 4.85     & 5.183 & {}\\
        $b_d$(kpc)    & 0.305     & 0.296 & {} \\
        $a_h$(kpc)    & 200     & 199.14 & {}\\
        $\chi^2$ & --   &  1.03  & {} \\ \hline \hline
     Constraints & Observed  & Best Fit & {}\\ \hline
        $V_r$    & see \cite{vc} & see Fig.\ref{fig:vc}& {}\\
        $\rho_{\odot}$ & 0.102$\pm$0.01 & 0.129 & {} \\
        $\sum_{1.1}$  & 74$\pm$6        & 69.0 & {}\\ \hline
    \end{tabular}
    \caption[Parameters of potential Model  II]{Parameters of Model II. $^*$ The values are extracted from Table 2 of \cite{Irrgang}. The best fit values of the parameters are obtained via $\chi^2$ minimisation. }
    \label{tab:model02}
\end{table*}
\begin{table*}
    \centering
    \begin{tabular}{|cccc|}
    \hline 
      Parameters & Value$^*$     & Best fit & Derived \\ \hline
        $M_b(10^{10}\text{M}_{\odot})$     & 1.020675  & 0.93 & 0.93\\
        $M_d(10^{10}\text{M}_{\odot})$    & 7.1982    & 8.88  & 8.714\\
        $M_h(10^{10}\text{M}_{\odot})$    & 330.615   & 161.2 &  97.789 \\ 
        $b_b$(kpc)    & 0.236     & 0.238 & {}\\
        $a_d$(kpc)    & 3.262     & 3.712 & {} \\
        $b_d$(kpc)    & 0.289     & 0.241 & {}\\
        $a_h$(kpc)    & 45.02     & 35.164 & {}\\  
        $\chi^2$ & --   &  1.03  & {} \\ \hline \hline
     Constraints & Observed  & Best Fit &{}  \\ \hline
        $V_r$    & see \cite{vc} & see Fig.\ref{fig:vc}&{}\\
        $\rho_{\odot}$ & 0.102$\pm$0.01 & 0.15 &{}\\
        $\sum_{1.1}$  & 74$\pm$6        & 68.28 &{}\\  \hline
        
    \end{tabular}
    \caption[Parameters of potential Model  III]{Parameters of Model III. $^*$ The values are extracted from Table 3 of \cite{Irrgang}. The best fit values of the parameters are obtained via $\chi^2$ minimization. }
    \label{tab:model03}
\end{table*}

\begin{table*}
    \centering
    \setlength\tabcolsep{2pt}
    \begin{tabular}{|c|c|c|c|c|c|c|}\hline
     
    Outputs & $R_{peri}$(kpc) & $R_{apo}$(kpc) & $Z_{max}$(kpc) & e & E (kpc/Myr)$^2$ & $L_z$(kpc$^2$/Myr) \\ \hline
    $\Omega$=41 kpc$^{-1}$ km/s  & 3.59$\pm$0.8 & 14.98$\pm$3.53 & 6.3$\pm$2.06 & 0.6$\pm$0.02 & -0.14$\pm$0.014 & -1.27$\pm$0.3 \\ \hline 
    $\Omega$=50 kpc$^{-1}$ km/s & 3.63$\pm$0.83 & 15.14$\pm$3.61 & 6.36$\pm$2.15 & 0.61$\pm$0.01 & -0.14$\pm$0.01 & -1.24$\pm$0.29\\ \hline
    $\Omega$=60 kpc$^{-1}$ km/s &  3.5$\pm$0.8 & 14.5$\pm$3.37 & 6$\pm$1.99 & 0.61$\pm$0.02 & -0.14$\pm$0.01 & -1.29$\pm$0.3 \\ \hline
    \end{tabular}
    \caption[Orbital parameters of three pattern speeds]{Orbital parameters derived with three pattern speeds of bar potential. Different pattern speeds lead to similar orbital parameters, suggesting that the influence of bar potential is weak, in agreement with the fact that FSR 1758 did not enter the bar region. }
    \label{tab:orbits_bar}
\end{table*}


\section{Results and Discussion}

For the sake of comparison, we remind the reader  that the orbit of FSR 1758 was recently studied by \citet{Simpson}. 
Adopting as  input initial conditions ($\alpha$, $\delta$, d$_{\odot}$, $\mu_{\alpha}$, $\mu_{\delta}$, $v_r$) = (262.806$^{\text{o}}$, -39.822$^{\text{o}}$, 11.5$\pm$ 1.0 kpc, -2.85$\pm$ 0.1 mas/yr, 2.55$\pm$ 0.1 mas/yr, 227$\pm$ 1 km/s) he found that  FSR 1758 possesses a retrograde orbit, with $R_{peri}=$ 3.8$\pm$0.9 kpc $R_{apo}=$ 16$^{+8}_{-5}$ kpc, and e=0.62$^{+0.05}_{-0.04}$.

\begin{figure*}
    \centering
    \includegraphics[width=\textwidth]{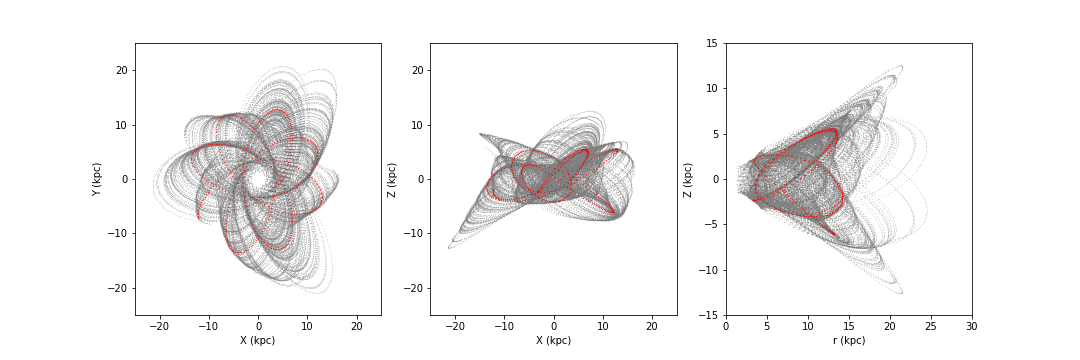}
    \includegraphics[width=\textwidth]{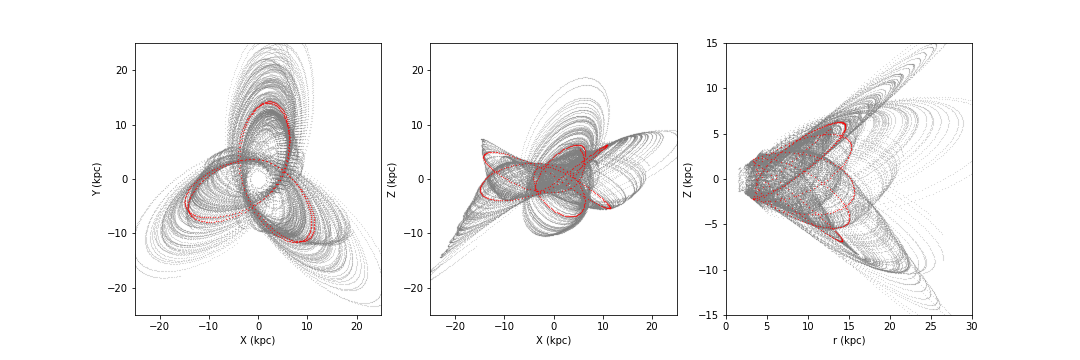}
    \includegraphics[width=\textwidth]{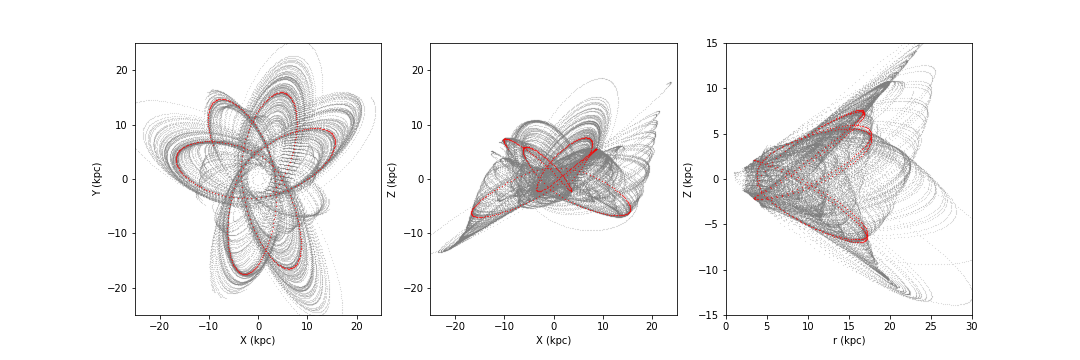}
    \caption[Orbits of FSR 1758  with error distribution from model II]{Orbits of FSR 1758 with error distributions obtained from Monte  Carlo Simulations. From top to bottom are orbits calculated from model I to model III including the bar potential. The red lines are orbits for the mean values of the phase space coordinates. The grey lines show 100 orbits randomly sampling the error distributions of the input conditions. }
    \label{fig:orbits}
\end{figure*}

Our results are shown in Tab.\ref{tab:outputs} and Tab.\ref{tab:orbits_bar}  for potentials without bar , and with bar component of different pattern speed, respectively.  The orbits of FSR 1758 including the  bar component with $\Omega=41$ kpc$^{-1} $km/s, are plotted in Fig.\ref{fig:orbits} and have the following features. In general,  $R_{peri}$ falls outside the bulge, the $R_{apo}$ locates far away from the center and $Z_{max}$ goes beyond the widely accepted height of Galactic thick disk. In addition, the orbit exhibits a high eccentricity of $\sim$ 0.6 and the cluster shows retrograde motions. We then confirm \citet{Simpson} basic results.

Often in the literature stars or clusters with retrograde motion or high eccentricity  are  considered  to origin from some accretion event. Hence, in order to investigate this association,  we constructed plots of of $L_z$ in semi-logarithm scale versus eccentricity by taking advantage of orbital parameters from \cite{ocs} for 488 OCs and \cite{Baumgardt} for GCs ( with two exceptions Ter.10 and Djor.1 from \citealt{Intruders}).\\

\noindent
GCs are grouped into: 
\begin{enumerate}

    \item possible accreted GCs from: 
    \begin{itemize}
        \item Gaia Sausage (\citealt{Myeong2018} and \citealt{Myeong2019}),
        \item Sagittarius GCs \citep{Forbes2010},
        \item Sequoia \citep{Myeong2019}, 
        \item Kraken \citep{Kruijssen2019},  and 
        \item Gaia Enceladus \citep{Helmi}.
    \end{itemize}
    \item in-situ GCs:
    \begin{itemize}
        \item bulge GCs listed in the Table 1. of \citealt{Bica},
        \item probable intruders  listed in the Table 2. of \citealt{Bica}, and 
        \item lastly, GCs not belong to any of them are grouped into halo GCs. 
    \end{itemize}
\end{enumerate}

\begin{figure*}
    \centering
    \includegraphics[width=.6\textwidth]{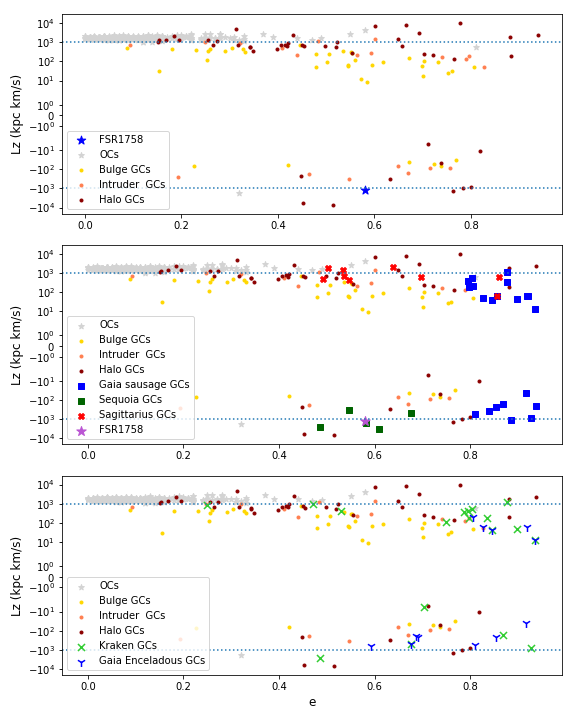}
    \caption[Lz-e plots]{Relationships between the angular momentum along z direction $L_z$ versus eccentricity e for the GC groups. Top panel: In-situ GCs associated with the Galactic halo, bulge or intruders into the bulge in comparison with OCs. Middle panel: GCs from three accretion events, Gaia Sausage, Sequoia and Sagittarius, compared with in-situ GCs and OCs. Bottom panel:  GCs from accretion the events Kraken and Gaia Enceladus, compared with in-situ GCs and OCs. }
    \label{fig:lz_e}
    
\end{figure*}

The plots are shown in Fig.\ref{fig:lz_e} and GCs assigned to each group are also listed in Table \ref{tab:GCs}.

It is important to underline that these plots show quite some differentiation among the various GC groups, which can then lead to very different interpretations. 
First, the top panel compares OCs with in-situ GCs. Because most OCs formed in the disk and rotate about the Galaxy in nearly circular clockwise orbits, they concentrate at high $L_z$ ($\geq 10^3$) and low eccentricity. However, unlike OCs, GCs in general are more dispersed in $L_z$ and e  and occupy both prograde and retrograde orbits.  Bulge GCs have $L_z< 10^3$. On the other hand, halo GCs have  higher $L_z$ in average, yet with a broader distribution. The lower $L_z$ of bulge GCs than the halo GCs agrees with a dissipative collapse formation scenario of the bulge, during which low angular momentum gas collapsed towards the inner parts of the Galaxy.  Possible intruder GCs seem to coincide with the  previous two groups. The values of the mean $L_z$ and standard deviation for the three groups are summarised in Tab.\ref{tab:in-situ}.

\begin{table}
\centering

\begin{tabular}{|c|c|c|c|c|}
\hline
\multirow{2}{*}{In-situ}  & \multicolumn{2}{c|}{Prograde}  & \multicolumn{2}{c|}{Retrograde} \\ \cline{2-5} 
                  & $<L_z>$     & $\sigma_{L_z}$   & \textless{}$L_z$\textgreater{} & $\sigma_{L_z}$ \\ \hline
OCs             &    1721.84    &    311.32        &     --     &     --         \\ \hline
Halo            &    1673.87    &   2169.60       &     -1675.05       &    2447.72    \\ \hline
Bulge           &   205.52      &    172.15      &       -54.12         &      13.56    \\ \hline
Intruders       &  538.29       &     441.77       &      -165.84       &    90.65      \\ \hline
\end{tabular}
\caption[Lz of in-situ clusters]{The mean angular momentum along z-axis and standard deviation of four groups of GCs formed in-situ. Units of $L_z$ and $\sigma_{L_z}$ are in kpc km/s. }
\label{tab:in-situ}
\end{table}

To continue our analysis of of Fig.\ref{fig:lz_e}, we now turn on the  middle and bottom panels  which show different grouping of GCs from different accretion events  as proposed by several authors. Most of the accreted GCs have eccentricity higher than 0.5. In particular, Gaia Sausage, Sequoia, and Sagittarius GCs in the middle panel show sharp grouping features in eccentricity:  e = 0.4-0.6 for the first two,  and 0.8-1.0 for the last one. In addition, apart from Gaia Sausage group which has a ratio of prograde/retrograde orbits 13/8, Sagittarius and Sequoia groups hold 100\% prograde and 100\% retrograde motion,  respectively. However, if we turn our attention on Kraken and Gaia Enceladus, GCs in each of this events do not  seem to relate to each other significantly, with prograde and retrograde orbits mixed together in a broad range of eccentricity. Some GCs even overlap between the two accretion events.  Most Gaia Enceladus GCs exhibit similar distribution as Gaia Sausage  GCs, in nice agreement with \cite{Piatti} who probed accretion events using the inclination and the eccentricity of GCs. 

As a consequence, we stress that GCs with retrograde or high eccentricity orbits are not special cases at all. For a system with high random motions as the Galactic halo, the mixing of prograde and retrograde orbits and the presence of high eccentricity orbits is quit expected. Accreted GCs should possess similar kinematic behaviour among them since they come from the same structure (stream, defunct dwarf galaxy, etc) which gravitationally bound them together, and therefore when the merging happened their collective behaviour should have been preserved due to the low density of the outer halo. The real question we are trying to answer here is  whether these collective behaviours  are distinct enough from the Milky Way GC properties to be solid indications accretion events.

Going back to our target, from the first and second plots of Fig.\ref{fig:lz_e}, FSR 1758 locates in the range of both Sequoia group and among {\it in-situ} halo GCs.
\cite{Myeong2019} has suggested FSR 1758 to be one member of Sequoia based on the distribution in action spaces. Here, though there are only five GCs in the Sequoia event, the probability that the clustering is a signature of accretion events cannot  be easily ruled out. Hence, FSR 1758 could be one of GCs in Sequoia dwarf galaxy accreted to the Milky Way. Similarly, because there are only five GCs, this evidence is not strong enough to support the clustering behaviour. An {\it in-situ} origin for the cluster is equally possible. 

In conclusion, basing only on kinematics, the origin of FSR 1758 cannot be distinguished among the two different scenarios:  it can equally be either an outer  halo intruder or an accreted cluster member of the Sequoia event. 
However, just recently \cite{Villanova} analysed  chemical components of FSR 1758 in detail by using high dispersion spectra for 9 stars and discovered Na-O anti-correlation in a  metal-poor GC, quite common among Galactic GCs . According to this study, the two components fit the mean Na and O abundance of other halo GCs very well when the second generation stars in them are excluded.  When considering all stars, depletion in O and enhancement in Na comes out. Apart from this, its $\alpha$ elements display the same trend with Galactic GCs as well as halo and thick disk stars and the trend is not commonly seen in extra-galactic objects. As a consequence, it would seem that FSR 1758 is more similar to {\it in-situ} halo GCs as far as chemistry is concerned.

\begin{table*}
\centering
\footnotesize
    \setlength\tabcolsep{2pt}
\begin{tabular}{|c|c|c|}
\hline
Accretion Events & Candidate GCs   & Sources                  \\ \hline
Gaia Sausage      & \begin{tabular}[c]{@{}l@{}}NGC 1851$^1$, NGC 1904$^1$,  NGC 2298$^1$, NGC 2808$^1$, \\ 
                                                NGC 5286$^1$, NGC 6779$^1$, NGC 6864$^1$, NGC 7089$^1$, \\ 
                                                NGC 362$^1$  , NGC 1261$^1$, NGC 4147$^2$, NGC 4833$^2$, \\ 
                                                NGC 5694$^2$, NGC 6544$^2$, NGC 6584$^2$, NGC 6712$^2$,\\ 
                                                NGC 6934$^2$, NGC 6981$^2$, NGC 7006$^2$, Pal 14$^2$, \\ 
                                                Pal 15$^2$.\end{tabular}
                    & \begin{tabular}[c]{@{}l@{}}$^1$\cite{Myeong2018} ,\\                                           $^2$\cite{Myeong2019}\end{tabular}\\ \hline
Sequoia   & \begin{tabular}[c]{@{}l@{}}FSR 1758, NGC 3201, NGC 5139, NGC 6101,\\ 
                                        NGC 5635, NGC 6388\end{tabular}                                 
            & \cite{Myeong2019}     \\ \hline
Sagittarius  & \begin{tabular}[c]{@{}l@{}}Ter 7, Arp 2, Ter 8, NGC 6715, NGC 4147,\\ 
                                            NGC 5634, Pal 12, AM 4, Whiting 1.\end{tabular}  
            & \cite{Forbes2010}                   \\ \hline
Kraken    & \begin{tabular}[c]{@{}l@{}}NGC 362, NGC 1261, NGC 3201, NGC 5139, \\ 
                                        NGC 5272, NGC 5897, NGC 5904, NGC 5946, \\ 
                                        NGC 6121, NGC 6284, NGC 6544, NGC 6584, \\ 
                                        NGC 6752, NGC 6864, NGC 6934.\end{tabular} 
        & \cite{Kruijssen2019}             \\ \hline
 \hline
Gaia Enceladus    & \begin{tabular}[c]{@{}l@{}}NGC 288, NGC 362, NGC 1851, NGC 1904, \\ 
                                                NGC 2298, NGC 4833, NGC 5139, NGC 6205, \\ 
                                                NGC 6341, NGC 6779, NGC 7089, NGC 7099.\end{tabular}        
                & \cite{Helmi}          \\ \hline \hline
 
In- Situ         & Candidate GCs      & Sources      \\ \hline
Bulge  & \begin{tabular}[c]{@{}l@{}}Ter 3, ESO 452-SC11, NGC 6256, NGC 6266, \\ 
                                    NGC 6304, NGC 6316, NGC 6325, NGC 6342,\\ 
                                    NGC 6355, Ter 2, Ter 4, HP 1, Lil 1, Ter 1, \\ 
                                    Ton 2, NGC 6401, Pal 6, Ter 5, NGC 6440, \\ 
                                    Ter 6, UKS 1, Ter 9, Djor 2, NGC 6522, \\ 
                                    NGC 6528, NGC 6539, NGC 6540, NGC 6553, \\ 
                                    NGC 6558, NGC 6569, BH 261, Mercer 5, \\ 
                                    NGC 6624, NGC 6626, NGC 6638, NGC 6637,\\ 
                                    NGC 6642, NGC 6652, NGC 6717, NGC 6723\end{tabular}                      
        & Table 1 of \cite{Bica}.\\ \hline

Intruders   & \begin{tabular}[c]{@{}l@{}}Lynga 7, NGC 6144, NGC 6171, NGC 6235, \\
                                        NGC6273,NGC 6287, NGC 6293, NGC 6352, \\ 
                                        NGC 6380, NGC 6388, NGC 6402, NGC 6441, \\
                                        NGC 6496, NGC 6517, NGC 6544, 2MS 2, \\
                                        IC 1276, Ter 12, NGC 6712.\end{tabular}                              
            & Table 2 of \cite{Bica} \\ \hline
            
Halo  & \begin{tabular}[c]{@{}l@{}}NGC 104, AM 1, Eridanus, Pal 2, NGC 2419, \\ 
                                    Pyxis, E 3, Pal 3, Pal 4, Crater, NGC 4372, \\ 
                                    NGC 5024, NGC 5053, NGC 5466, NGC 5824,\\ 
                                    Pal 5, NGC 5927, NGC 5986, FSR 1716, \\ 
                                    NGC 6093, NGC 6139, NGC 6229, NGC 6218, \\ 
                                    FSR 1735, NGC 6254, NGC 6333, NGC 6356, \\ 
                                    IC 1257, NGC 6366, NGC 6362, NGC 6397,\\ 
                                    NGC 6426, Djor 1$^*$, Ter 10$^*$, NGC 6535, \\ 
                                    NGC6541, ESO 280, Pal 8, NGC 6656, \\ 
                                    NGC 6681, NGC 6749, NGC 6760, Pal 10, \\ 
                                    NGC 6809, Pal 11, NGC 6836, Pal 13, NGC 7492.\\
                                    \end{tabular} 
& \multicolumn{1}{p{3.5cm}|}{\vspace{-2cm}GCs in \cite{Baumgardt} not listed in any of previous group. $^*$ Orbital parameters of Djorgoski 1 and Terzan 10 are taken from \cite{Intruders} since it has more reliable distance measurements than \cite{Baumgardt} in which Terzan 10 was revealed to be a bulge cluster. However, according to \cite{Intruders} it's a halo intruder.} \\ \hline
\end{tabular}
\caption[The {\it in-situ} and accreted GCs]{Lists of accreted and in-situ GCs categorised into different groups.}
\label{tab:GCs}
\end{table*}

\section{Conclusions}

In this study the orbits of FSR 1758 were derived employing three different galactic potential models by integrating the equation of motion backward in time  for 1.25 Gyr with an efficient and precise algorithm, namely the Gauss-Radau spacings of $15^{th}$ order. The resulting orbits are in nice agreement with the one in \cite{Simpson}, having orbital parameters  3 kpc $<R_{peri}<$ 4 kpc,  14 kpc $<R_{apo}<$ 16 kpc,  $Z_{max}\sim 6$ kpc and e $\sim$ 0.6. Furthermore, a potential with bar component was then added to observe its influence  on the cluster orbit. It was assumed that the mass of the bar is transferred from the mass of the bulge instantly. When viewed in the bar-axis reference, the cluster never enters the inner region of the bar, matching the fact that there are no significant changes in orbital parameters.
    
Because its apo-galactic distance is far away from the Galactic centre and the maximum height of its orbit exceeds the height of Galactic thick disk, FSR 1758 has to be considered an intruder from the outer Galactic halo. However, whether it is an {\it in-situ} GC that formed inside the Galactic halo or belongs to one of  accreted GCs left over after the merger of satellite dwarf galaxies into the Milky Way is hard to say. FSR 1758 possesses a retrograde orbit with high eccentricity that are thought to be signatures of accretion events, but not uncommon among {\it in  situ} Galactic halo globulars. 

\noindent
(Fig.\ref{fig:lz_e}) shows that in the cases of Sagittarius, Gaia Sausage, and Sequoia, globulars clusters possess  a narrow $L_z$ distribution and high eccentricity. Besides,  retrograde orbits are not a distinctive feature of accretion events anymore. All GCs in Sagittarius have prograde orbits, while all GCs in Sequoia have retrograde orbits. What is even more surprising is that  Gaia Sausage harbours both prograde and retrograde orbits of very high eccentricity, in line with the statement that it is a head-on collision event. \\

In the cases of Kraken and Gaia Enceladus, they exhibit a significant spread in eccentricity,  but mostly higher than 0.5. Prograde and retrograde orbits are mixed together again. However, being clusters less concentrated here, they show some overlapping with other events, especially Gaia Sausage. This might suggest that different accretion groups may be in fact come from a unique large merging event, or simpler, that we do not have enough information to assign GCs  to a given parent populations firmly.\\
 
As for FSR 1758, it falls indeed in the region of Sequoia event. Although Sequoia GCs seem to be confined in a small region, there are only four candidates GCs in this event. As a consequence, 
we are not able to firmly confirm  that  FSR 1758 is really from Sequoia. It might equally be an  {\it in-situ} halo GCs, if we limit ourselves to orbital parameter analysis.
According to the detailed chemical abundance analysis done by \cite{Villanova}, FSR 1758 is  found to contain similar trend in $\alpha$ elements and Na-O anti-correlation for metal-poor halo GCs. 
This, in our opinion, lends more support to  a more conservative scenario in which FSR 1758 is most probably a halo GC formed inside the Milky Way .

\begin{acknowledgements}
 Fu Chi Ye acknowledges the European Union founded Astromundus program ({\tt https://www.uibk.ac.at/astromundus/}).
 Vladimir Korchagin acknowledges the financial support from grant No 18-12-00213
of Russian science foundation. The comments of an anonymous referee are greatly appreciated.
\end{acknowledgements}

\bibliography{references}
\bibliographystyle{aa}
\end{document}